\documentclass[11pt]{article}

\usepackage[utf8]{inputenc}
\usepackage[T1]{fontenc}
\usepackage{lmodern}

\usepackage[a4paper,margin=1in]{geometry}
\usepackage{microtype}

\usepackage{amsmath,amssymb,amsfonts}
\usepackage{graphicx}
\usepackage{booktabs}
\usepackage{physics}

\usepackage[hidelinks]{hyperref}
\usepackage{url}

\title{Landauer cost in a continuous vacuum/no-vacuum measurement}

\author{
	Lorenzo Pirovano\thanks{Email: \texttt{lorenzopirovano.res@gmail.com}}
}

\date{December 27, 2025} 

\begin{document}
	\maketitle
	
	\begin{abstract}
		We study the thermodynamic cost associated with maintaining a continuous binary record of a \emph{vacuum/no-vacuum} measurement. For a single bosonic mode we model the operational vacuum test as a coarse-grained Poissonian click/no-click process with measurement strength \(\gamma\) and time resolution \(\tau\), where the click probability in each bin depends on the instantaneous vacuum occupation \(p_0(t)\). Treating the measurement outcomes as a classical register that is reset after every bin, Landauer's principle yields a lower bound on the average dissipated heat per bin proportional to the Shannon entropy of the record, and hence a minimal dissipation power. In the high time-resolution limit \(\tau\to 0\) at fixed click rate, this power bound diverges only logarithmically, making explicit the need for an operational bandwidth/coarse-graining cutoff. We then extend the framework to \(N\) monitored modes in a region, emphasizing that the relevant quantity is the joint entropy \(H(\mathbf{Y}_t)\) of the multi-bit record, which reduces to \(N H_2(P_\tau)\) for independent outcomes and can be smaller in the presence of correlations due to compressibility. We discuss practical parameter estimates for circuit-QED photon monitoring and present a speculative horizon-based thought experiment to illustrate how an entropy budget and a coarse-graining timescale constrain the maximal information throughput and its associated Landauer cost.
	\end{abstract}


	\section{Introduction}
	The act of measurement is inseparable from the production of information.
	In modern platforms---from quantum optics to superconducting circuits---it is routine to continuously monitor an output channel and to maintain a time-stamped classical record of ``events'' (clicks) or continuous readouts.
	Such measurement records underpin quantum trajectories, real-time feedback, and error correction, and they can be processed into compact sufficient statistics or used directly as raw bit streams \cite{WisemanMilburn2010,Jacobs2014,PlenioKnight1998,Carmichael1993}.
	
	A complementary viewpoint, developed in information thermodynamics, is that classical information is a physical resource whose manipulation has an unavoidable thermodynamic footprint.
	In particular, Landauer's principle states that any logically irreversible operation---paradigmatically, the erasure/reset of a memory register---requires the dissipation of heat at least of order $k_B T$ times the entropy removed from the memory \cite{Landauer1961,Bennett1982,Bennett2003,ReebWolf2014}.
	This link has been confirmed experimentally in mesoscopic one-bit memories \cite{Berut2012} and has become part of the broader framework connecting entropy production, information, and feedback \cite{Parrondo2015,Goold2016}.
	Within this perspective, a continuously operating detector that produces classical outcomes at a fixed bandwidth must either (i) accumulate an ever-growing memory, or (ii) regularly reset/erase its register to sustain stationary operation.
	In case (ii), Landauer's principle yields an \emph{operational} lower bound on the average dissipated heat per unit time, controlled by the information rate of the measurement record.
	
	In this paper we apply this logic to a particularly coarse-grained monitoring task: a continuous \emph{vacuum/no-vacuum} test.
	Given a bosonic mode, the operational question is not ``how many excitations are present?'', but whether the state lies in the vacuum subspace or in its orthogonal complement.
	This type of thresholding arises naturally whenever a detector is sensitive only to the presence of \emph{any} excitation above the ground state, and it is closely related to photon-detection primitives used in circuit-QED and microwave quantum optics \cite{Blais2021,Lescanne2019,Royer2018}.
	We model the measurement as a click/no-click process with finite time resolution $\tau$ (equivalently, finite bandwidth), and we treat the time-binned outcomes as a classical register that is reset after every bin.
	
	Our main message is simple: once the classical record is taken seriously as a physical memory that is periodically erased, the thermodynamic cost is governed not by the energy of the measured mode, but by the \emph{Shannon entropy} of the record.
	For a single monitored mode, the record in each bin is a biased bit whose entropy is the binary Shannon entropy $H_2(P_\tau)$ (in nats), where $P_\tau$ is the click probability in a bin.
	Landauer's principle then yields a minimal heat cost per bin $Q_{\min}\ge k_B T\,H_2(P_\tau)$ and, dividing by $\tau$, a minimal dissipation power that scales as $(k_B T/\tau)H_2(P_\tau)$.
	Because near-vacuum states produce strongly biased outcomes, the record is highly compressible and the optimal erasure cost can be parametrically smaller than the worst-case $k_B T\ln 2$ per bin.
	
	A key technical point is the role of coarse-graining.
	As $\tau\to 0$ at fixed instantaneous click rate, the \emph{power} bound grows without bound, but only logarithmically in $1/\tau$.
	This makes explicit that an operational cutoff---set by detector bandwidth, sampling time, or an explicit coarse-graining protocol---is not a mere technicality but is required for any finite information-throughput (and hence finite Landauer cost) statement.
	We then extend the discussion to $N$ monitored modes in a region, emphasizing that the relevant information-theoretic quantity is the \emph{joint} entropy $H(\mathbf{Y}_t)$ of the $N$-bit record per bin.
	For independent outcomes this reduces to $N H_2(P_\tau)$, but correlations can reduce the cost via redundancy and compressibility (equivalently, a smaller entropy rate).
	
	Finally, we connect the phenomenological model to circuit-QED parameter scales and sketch a speculative horizon-based thought experiment (in the spirit of holographic bookkeeping) to illustrate how an entropy budget and a coarse-graining timescale constrain the maximal information throughput and the associated Landauer cost \cite{GibbonsHawking1977,Bekenstein1973,Hawking1975,tHooft1993,Susskind1995,Bousso2002}.
	
	\section{Setup: a single-mode continuous vacuum test}
	We consider one bosonic mode with Hamiltonian
	\begin{equation}
		H = \hbar \omega \left(a^\dagger a + \frac{1}{2}\right)
	\end{equation}
	We are interested in testing for the vacuum occupation probability $p_0$ versus its complement $p_\mathrm{n0} = 1 - p_0$. To do so we define the vacuum projector $\Pi_0 = \ket{0} \bra{0}$.
	If we allow for a time-dependent state $\rho(t)$, we have
	\begin{equation}
		p_0(t) = \Tr [\rho(t) \Pi_0] 
	\end{equation}
	which is an expression for the instantaneous vacuum occupation probability. We note that in general the bosonic mode is just an example, as $p_0$ is usually defined for every kind of mode and the test compresses all the excitations above the ground state to a binary outcome $p_0$ versus $p_\mathrm{n0}$.
	\par The test we are referring to is a simple operational \emph{vacuum test}: a binary measurement attempting to decide whether the state lies	in $\ket{0}$ or in the orthogonal subspace. We imagine this test as a click/no-click probe where a no-click means that we find the ground state.
	
	\paragraph*{Test operator and record entropy.} To model the test, we use a phenomenological approach and write a coarse-grained binary measure, a jump operator \cite{Carmichael1993,Dalibard1992,PlenioKnight1998} that monitors the click case as a Poisson event, hence acts on the non-vacuum subspace:
	\begin{equation}
		J = \sqrt{\gamma} \left(\mathbb{I} - \Pi_0\right)
	\end{equation}
	where $\gamma$ represents the measurement strength. The Poisson instantaneous click rate is therefore
	\begin{equation}
		\lambda(t) = \Tr [\rho(t) J^\dagger J] = \gamma \left(1-p_0(t)\right)
	\end{equation}
	If we measure the click rate over a coarse-grained time bin of duration $\Delta t = \tau$, that renders a finite time resolution or a finite detector bandwidth, the click probability in this Poisson process is
	\begin{equation}
		P_\tau(t) = 1-e^{-\lambda(t)\tau} = 1-e^{-\gamma \left(1-p_0(t)\right)\tau}
	\end{equation}
	Also, for each time bin, the result is recorded into a classical register $Y \in \lbrace \text{click,no-click}\rbrace$ with probabilities $\lbrace P_\tau,1-P_\tau\rbrace$. The Shannon entropy \cite{Shannon1948} per bin (in nats) is
	\begin{equation}
		H_2(P_\tau)= -P_\tau\ln P_\tau-(1-P_\tau)\ln(1-P_\tau)
	\end{equation}
	The maximum $H_2=\ln 2$ occurs at $P_\tau=1/2$, i.e.
	\begin{equation}
		P_\tau=\frac12 \quad\Longleftrightarrow\quad \gamma(1-p_0)\tau=\ln 2.
		\label{eq:maxinfo_condition}
	\end{equation}
	This condition shows that vacuum-dominated states ($p_0\approx 1$) naturally yield $P_\tau\ll1$ and hence a strongly biased record with $H_2\ll\ln2$ unless $\gamma\tau$ is correspondingly large.
	
	\paragraph*{Remark (effective hazard model).}
	The bin probability $P_\tau(t)=1-e^{-\lambda(t)\tau}$ should be read as an \emph{effective hazard-rate parametrization} for the statistics of the time-binned record, with $\lambda(t)=\gamma(1-p_0(t))$ frozen at the bin start.
	We do not specify a full bin-wise measurement instrument nor the intra-bin conditional update.
	In particular, for $\gamma\tau\gtrsim 1$ this effective model can yield $P_\tau(t)>1-p_0(t)$, so it should not be interpreted as a literal single-shot threshold POVM acting on a perfectly frozen state.
	A strictly bin-wise threshold model with frozen $p_0$ would instead give
	$P_\tau^{\mathrm{(thr)}}(t)=(1-p_0(t))(1-e^{-\gamma\tau})\le 1-p_0(t)$, and both expressions coincide to leading order as $\tau\to0$.
	
	\paragraph*{Landauer bound per bin.} We suppose now that, surrounding the system, we have an effective bath or thermal reservoir at temperature $T$, that can also be an effective temperature determined by the measurement context. In presence of a continuous vacuum test where records of the test are reset at each time bin, the minimal average heat dissipated per bin is bounded by Landauer \cite{Landauer1961}:
	\begin{equation}
		Q_{\min}^{(1)}(t)\ \ge\ k_B T\,H_2(P_\tau(t)).
		\label{eq:landauer_perbin}
	\end{equation}
	Dividing by the bin duration gives a minimal dissipation power for one monitored mode:
	\begin{equation}
		\dot Q_{\min}^{(1)}(t)\ \ge\ \frac{k_B T}{\tau}\,H_2 \Big(1-e^{-\gamma(1-p_0(t))\tau}\Big).
		\label{eq:power_single_mode}
	\end{equation}
	Equation \eqref{eq:power_single_mode} is an \emph{operational} lower bound: it applies to the thermodynamic cost of maintaining a stream of classical records that are reset at every time bin. It is not a statement about the absolute energy of the mode itself, and it vanishes in the limit $p_0\to 1$ (perfect vacuum) at fixed $\gamma\tau$ because the record becomes deterministic.
	
	\par Equation \eqref{eq:power_single_mode} assumes an \emph{optimal} reset/erasure protocol that can exploit the bias of the outcomes.
	If the readout/reset mechanism is instead fixed or inefficient (e.g., it effectively treats each bin as a full unbiased bit), the actual dissipation can be larger, up to the worst-case value
	\begin{equation}
		\dot Q^{(1)}(t)\ \approx \frac{k_B T}{\tau}\,\ln 2 .
	\end{equation}
	Note that this worst-case scaling applies to any binary record and is not specific to the vacuum test considered here.
	
	\paragraph*{Small-bin limit $\tau\to 0$ and logarithmic divergence of the power bound.}
	In the high time-resolution (small-bin) regime one has $P_\tau = 1-e^{-\lambda\tau}\simeq \lambda\tau$ for $\lambda\tau\ll 1$, with $\lambda=\gamma(1-p_0)$.
	Expanding the binary Shannon entropy for small $P$ gives
	\begin{equation}
		H_2(P)= -P\ln P-(1-P)\ln(1-P)\;=\;P\big[1-\ln P\big]+O(P^2),
	\end{equation}
	so that
	\begin{equation}
		H_2(P_\tau)\simeq (\lambda\tau)\,\big[1-\ln(\lambda\tau)\big].
	\end{equation}
	Substituting into the Landauer power bound \eqref{eq:power_single_mode} yields
	\begin{equation}
		\dot Q_{\min}^{(1)}(t)\ \ge\ \frac{k_B T}{\tau}\,H_2(P_\tau)
		\ \simeq\ k_B T\,\lambda(t)\,\big[1-\ln(\lambda(t)\tau)\big].
	\end{equation}
	Therefore, as $\tau\to 0$ at fixed instantaneous rate $\lambda(t)$, the bound grows without limit, but only \emph{logarithmically} in $1/\tau$.
	This reflects the fact that arbitrarily fine time binning increases the information throughput of the classical record, and it makes explicit the need for an operational cutoff (finite detector bandwidth or finite coarse-graining time) for the bound to remain finite and physically meaningful.
	
	\begin{figure}
		\centering
		\includegraphics[width=0.7\linewidth]{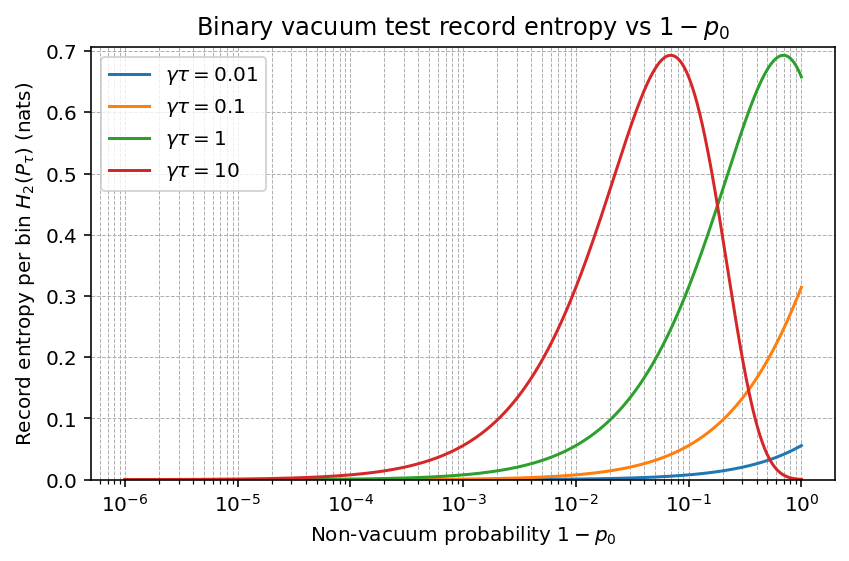}
		\caption{Record entropy per bin as a function of non-vacuum probability, once the product $\gamma \tau$ is fixed.}
		\label{fig:entropy}
	\end{figure}

	\section{Testing on many modes in a region}
	Until now we calculated the thermodynamic cost of testing the vacuum only on one mode. The goal of this section is to evaluate a total cost of a test done on all the modes contained in a given region. Defining a finite set of modes associated with a spatial region requires an explicit cutoff (e.g., detector bandwidth or spatial resolution); otherwise the number of modes diverges.
	
	\par For $N$ monitored modes (labeled by $k$), the minimal dissipation power is additive under independent readout/reset,
	\begin{equation}
		\dot Q_{\min}^{\mathrm{Tot}}(t)\ \ge\ \frac{k_B T}{\tau}\sum_{k=1}^N
		H_2 \Big(1-e^{-\gamma_k(1-p_{0,k}(t))\tau}\Big).
	\end{equation}
	For identical modes, $\gamma_k=\gamma$ and $p_{0,k}(t)=p_0(t)$, this reduces to $N$ times the single-mode bound.
	
	\par Another way to look at the problem is to think about the maximum number of modes that I can count given the maximum capacity of available record. More generally, the Landauer cost is set by the Shannon entropy of the \emph{joint} classical record produced in each time bin.
	Let $\mathbf{Y}_t=(Y_{1,t},\dots,Y_{N,t})$ denote the $N$-bit click/no-click record collected over the bin $[t,t+\tau]$.
	If the register storing $\mathbf{Y}_t$ is reset after each bin, then an operational Landauer bound reads
	\begin{equation}
		Q_{\min}^{\mathrm{Tot}}(t)\ \ge\ k_B T\, H(\mathbf{Y}_t),
	\end{equation}
	where $H(\mathbf{Y}_t)$ is the Shannon entropy (in nats) of the joint distribution $P(\mathbf{Y}_t)$.
	\begin{equation}
		H(\mathbf{Y}_t)
		:= -\sum_{\mathbf{y}}
		P(\mathbf{Y}_t=\mathbf{y})\,\ln P(\mathbf{Y}_t=\mathbf{y}) \qquad \mathbf{y}\in\{\mathrm{click},\mathrm{no\text{-}click}\}^N
	\end{equation}
	In particular, if the $N$ outcomes are independent and identically distributed with $\Pr(Y_{k,t}=\text{click})=P_\tau(t)$ for all $k$,
	then $H(\mathbf{Y}_t)=\sum_{k=1}^N H(Y_{k,t}) = N H_2(P_\tau(t))$, and therefore
	\begin{equation}
		Q_{\min}^{\mathrm{Tot}}(t)\ \ge\ N\,k_B T\, H_2(P_\tau(t)),
		\qquad
		\dot Q_{\min}^{\mathrm{Tot}}(t)\ \ge\ \frac{N k_B T}{\tau}\, H_2(P_\tau(t)).
	\end{equation}
	If correlations are present across modes, one must use the joint entropy $H(\mathbf{Y}_t)$ (or equivalently the entropy rate under suitable stationarity assumptions), which can be strictly smaller than $N H_2(P_\tau(t))$ due to redundancy/compressibility of the record.
	
	\par Switching to densities in a volume $V$ the specification is straightforward
	\begin{equation}
		\frac{Q_{\min}^{\mathrm{Tot}}(t)}{V}\ \ge\ \frac{N}{V}\,k_B T\, H_2(P_\tau(t)),
		\qquad
		\frac{\dot Q_{\min}^{\mathrm{Tot}}(t)}{V}\ \ge\ \frac{N k_B T}{V \tau}\, H_2(P_\tau(t)).
	\end{equation}
	and again we require a form for $N/V$ which depends of a chosen field theory. We can make the choice depending on how model-dependent we want to be.
	
	\paragraph*{Method 1: Modes per volume as phenomenological parameter.} This is the simplest and least model-dependent choice: we simply put
	\begin{equation}
		\nu := \frac{N}{V}
	\end{equation} 
	as experimentally determined by the coarse-graining of the detector. In this case we simply make no choice.
	
	\paragraph*{Method 2: Spatial resolution and mode per voxel.}
	A simple operational cutoff is provided by a finite spatial resolution $\ell$.
	We partition the region into volumetric pixels (voxels) of linear size $\ell$, so that the number of resolvable cells is $V/\ell^3$ (up to boundary effects). If we associate $g$ effectively independent bosonic modes to each voxel (e.g., accounting for internal degeneracies), then
	\begin{equation}
		N \simeq g\,\frac{V}{\ell^3},
		\qquad\Longrightarrow\qquad
		\nu := \frac{N}{V} \simeq \frac{g}{\ell^3}.
		\label{eq:nu_voxel}
	\end{equation}
	Here $\ell$ should be interpreted as the detector coarse-graining length (e.g., pixel size or point-spread function scale). This prescription makes explicit that finer spatial resolution increases the operational mode density and hence the minimal dissipation density.
	
	\paragraph*{Method 3: Spectral cutoff and state density $D(\omega)$.}
	A complementary operational cutoff is spectral: we only monitor modes within a frequency band $\omega\in[\omega_{\min},\omega_{\max}]$ set by the detector bandwidth and/or by the chosen coarse-graining.
	Introducing the density of modes per unit volume and per unit angular frequency, $D(\omega)$, we can write
	\begin{equation}
		\nu := \frac{N}{V} = \int_{\omega_{\min}}^{\omega_{\max}} D(\omega)\,d\omega,
		\label{eq:nu_DoS_general}
	\end{equation}
	or, for a narrow band of width $\Delta\omega$ around $\omega_0$,
	\begin{equation}
		\nu \simeq D(\omega_0)\,\Delta\omega.
		\label{eq:nu_narrowband}
	\end{equation}
	The specific functional form of $D(\omega)$ depends on the physical realization of the bosonic modes.
	For instance, in a 3D continuum with approximately linear dispersion $\omega=v|k|$ one has
	\begin{equation}
		D(\omega)= g\,\frac{\omega^2}{2\pi^2 v^3}
		\qquad\Longrightarrow\qquad
		\nu = g\,\frac{\omega_{\max}^3-\omega_{\min}^3}{6\pi^2 v^3}.
		\label{eq:nu_linear_dispersion}
	\end{equation}
	In an operational setting, the spectral window $[\omega_{\min},\omega_{\max}]$ (or $\Delta\omega$)
	should be identified with the detector bandwidth; one may also relate it to the bin time $\tau$ through an effective resolution $\Delta\omega\sim 1/\tau$ as an order-of-magnitude estimate.

	\section{Application: continuous vacuum/no-vacuum monitoring in circuit QED}
	
	To connect the phenomenological vacuum test to an actual laboratory setting, consider a single electromagnetic mode (a microwave cavity or resonator mode) in a circuit-QED platform \cite{Blais2021}. The operational task is to produce, every coarse-grained time bin of duration	$\tau$, a binary classical record $Y_t\in\{\mathrm{click},\mathrm{no\text{-}click}\}$ indicating whether the
	mode is ``non-vacuum'' (at least one photon present) or ``vacuum''.
	A natural realization of the ``click'' event is the detection of photons leaking out of the cavity into a monitored transmission line \cite{Royer2018,Lescanne2019}. In that case, the instantaneous click rate is	set by the cavity energy-decay rate $\kappa$ (photon loss rate) and by the overall detection efficiency $\eta\in[0,1]$. A simple estimate is therefore
	\begin{equation}
		\gamma \simeq \eta\,\kappa,
	\end{equation}
	so that the coarse-grained click probability per bin becomes
	\begin{equation}
		P_\tau(t)=1-e^{-\gamma (1-p_0(t))\tau}
		\;\simeq\;
		1-e^{-\eta\kappa (1-p_0(t))\tau}.
	\end{equation}
	As an order-of-magnitude example, take a bin size $\tau=1~\mu\mathrm{s}$, a cavity decay rate $\kappa/2\pi\simeq 1~\mathrm{MHz}$ (hence $\kappa\simeq 6.3\times 10^6~\mathrm{s}^{-1}$), and a detection efficiency $\eta\simeq 0.8$. This yields an effective measurement strength
	\begin{equation}
		\gamma \simeq \eta\kappa \simeq 5\times 10^6~\mathrm{s}^{-1}.
	\end{equation}
	Assume the cavity is close to vacuum with $p_0=0.99$ (i.e.\ a non-vacuum probability
	$1-p_0=10^{-2}$). Then
	\begin{equation}
		P_\tau = 1-e^{-\gamma(1-p_0)\tau}
		=1-e^{-(5\times 10^6)(10^{-2})(10^{-6})}
		=1-e^{-0.05}
		\simeq 4.9\times 10^{-2}.
	\end{equation}
	The classical record is therefore strongly biased toward ``no-click'', and its Shannon entropy per bin is significantly below $\ln 2$:
	\begin{equation}
		H_2(P_\tau)\simeq 0.195~\text{nats} \;\;\;(\simeq 0.281~\text{bits}).
	\end{equation}
	If the record register is reset after each bin against a reservoir at temperature $T$, Landauer's	principle implies a minimal average heat cost per bin following Equation \ref{eq:landauer_perbin} and a minimal dissipation power when $\tau$ is taken into account.
	For the above numbers, one finds:
	\begin{align}
		T=20~\mathrm{mK}:&\qquad
		Q_{\min}^{(1)}\approx 5.4\times 10^{-26}~\mathrm{J/bin},\qquad
		\dot Q_{\min}^{(1)}\approx 5.4\times 10^{-20}~\mathrm{W},
		\\
		T=300~\mathrm{K}:&\qquad
		Q_{\min}^{(1)}\approx 8.1\times 10^{-22}~\mathrm{J/bin},\qquad
		\dot Q_{\min}^{(1)}\approx 8.1\times 10^{-16}~\mathrm{W}.
	\end{align}
	These values are far below typical electronics power budgets, but they provide an operational thermodynamic lower bound associated with maintaining and erasing the stream of classical outcomes.
	If $N$ independent modes are monitored in parallel with comparable parameters, the minimal cost is additive:
	\begin{equation}
		\dot Q_{\min}^{\mathrm{Tot}} \ \ge\ N\,\dot Q_{\min}^{(1)}
		\ =\ \frac{N k_B T}{\tau}\,H_2(P_\tau).
	\end{equation}
	Because $P_\tau\ll 1$ in the near-vacuum regime, the record is highly compressible. An optimal erasure protocol can exploit this bias, paying only $k_B T\,H_2(P_\tau)$ per bin on average. In contrast, a	non-adaptive or inefficient reset mechanism that effectively treats each bin as an unbiased bit incurs a larger cost up to the worst-case value $k_B T\ln 2$ per bin, i.e.\ larger by a factor $\ln 2/H_2(P_\tau)$ for the same physical measurement record.

	\section{Discussion: What if the region was the universe?}
	As an heuristic and speculative thought experiment, we could try to impose that the universe itself is continuously performing a binary test on its entire bulk region, storing the results on the horizon's area, following the ideas of the holographic principle. We could also imagine that the "ruler" used by the horizon is too wide to scout a single mode, hence we would end at maximum cost for the test at $H_2=\ln 2$ everywhere due to horizon-limited coarse-graining. The last hypothesis is to take the limit of a de Sitter horizon. \cite{GibbonsHawking1977,Bekenstein1973,Hawking1975,tHooft1993,Susskind1995,Bousso2002}
	
	\paragraph*{Cost for a single test on a single mode.}
	Assume the record is erased (or equivalently ``thermalized'') against a de Sitter horizon, so that the relevant temperature is the Gibbons--Hawking temperature
	\begin{equation}
		T_{\mathrm{dS}}=\frac{\hbar H}{2\pi k_B},
		\label{eq:T_dS}
	\end{equation}
	with $H$ the (asymptotic) Hubble rate of the de Sitter patch. In the maximally coarse-grained regime where each bin produces an unbiased outcome, $H_2=\ln 2$, Landauer's principle gives the minimal heat per bin for a single binary test as
	\begin{equation}
		Q_{\min}^{(1)} \ \ge\ k_B T_{\mathrm{dS}} \ln 2
		\;=\; \frac{\hbar H}{2\pi}\,\ln 2.
		\label{eq:Qbit_dS}
	\end{equation}
	Equivalently, each bin carries an entropy $\Delta S = k_B\ln 2$ that must be exported to (and can be absorbed by) the horizon degrees of freedom in a reversible idealization.
	
	\paragraph*{Cost for a single test on all modes.}
	If the bulk region contains $N$ effectively monitored bosonic modes and each mode yields (by assumption) an unbiased
	bit per bin, the joint record has entropy $H(\mathbf Y_t)=N\ln 2$ and the Landauer cost per bin scales as
	\begin{equation}
		Q_{\min}^{\mathrm{Tot}} \ \ge\ N\,k_B T_{\mathrm{dS}} \ln 2.
		\label{eq:Qtot_dS}
	\end{equation}
	If the horizon acts as the memory that ultimately stores this information, then the entropy dump per bin is
	\begin{equation}
		\Delta S_{\mathrm{hor}} = k_B H(\mathbf Y_t)= N k_B \ln 2.
	\end{equation}
	Using the Bekenstein--Hawking area law $S_{\mathrm{hor}}=k_B A/(4\ell_p^2)$, this corresponds to a minimal horizon-area
	increase
	\begin{equation}
		\Delta A = 4\ell_p^2\,N\ln 2,
		\qquad
		\ell_p^2=\frac{\hbar G}{c^3}.
		\label{eq:Area_per_bin}
	\end{equation}
	Moreover, the finite de Sitter entropy budget
	\begin{equation}
		S_{\mathrm{dS}}=\frac{k_B A_{\mathrm{dS}}}{4\ell_p^2},
		\qquad
		A_{\mathrm{dS}}=4\pi r_{\mathrm{dS}}^2,\ \ r_{\mathrm{dS}}=\frac{c}{H},
		\label{eq:SdS}
	\end{equation}
	implies a holographic upper bound on the number of unbiased bits that can be stored,
	\begin{equation}
		N \ \leq\ \frac{S_{\mathrm{dS}}}{k_B\ln 2}
		\;=\;\frac{A_{\mathrm{dS}}}{4\ell_p^2\ln 2}.
		\label{eq:Nmax_holo}
	\end{equation}
	If, in addition, one considers the maximal information throughput compatible with the de Sitter entropy budget (i.e., $N$ saturating \eqref{eq:Nmax_holo} over an effective reset), then the corresponding per-bin energy scale becomes
	\begin{equation}
		Q_{\min}^{\mathrm{Tot}} \bigg|_{N=S_{\mathrm{dS}}/(k_B \ln 2)} \ \ge T_{\mathrm{dS}} S_{\mathrm{dS}}
		\;=\;\frac{c^5}{2GH},
		\label{eq:TS_identity}
	\end{equation}
	which matches the characteristic energy associated with the vacuum (dark-energy) content inside the de Sitter static	patch. In terms of energy density, with $V = V_\mathrm{dS} = (4/3)\pi r_{\mathrm{dS}}^3 $:
	\begin{equation}
		\frac{Q_{\min}^{\mathrm{Tot}}}{V_\mathrm{dS}} \bigg|_{N=S_{\mathrm{dS}}/(k_B \ln 2)} \ \ge \frac{T_{\mathrm{dS}} S_{\mathrm{dS}}}{V_\mathrm{dS}}
		\;=\;\frac{\Lambda c^4}{8 \pi G} = \rho_\Lambda
	\end{equation}
	
	\paragraph*{Continuous test cost on all modes.}
	If the test is repeated every bin of duration $\tau$, the minimal dissipation power associated with resetting the
	record is
	\begin{equation}
		\dot Q_{\min}^{\mathrm{Tot}}
		\ \ge\ \frac{N k_B T_{\mathrm{dS}}}{\tau}\,\ln 2.
		\label{eq:power_dS_general}
	\end{equation}
	A natural coarse-graining timescale for horizon-limited readout is the horizon timescale $\tau\sim H^{-1}$
	(up to order-one factors). With this choice,
	\begin{equation}
		\dot Q_{\min}^{\mathrm{Tot}} \ \gtrsim\ N\,\frac{\hbar H^2}{2\pi}\,\ln 2.
		\label{eq:power_dS_H}
	\end{equation}
	If, again, the de Sitter entropy budget is saturated, and using again the horizon timescale, we have in terms of power:
	\begin{equation}
		\dot Q_{\min}^{\mathrm{Tot}} \bigg|_{N=S_{\mathrm{dS}}/(k_B \ln 2)}
		\ \ge\ \frac{S_{\mathrm{dS}} T_{\mathrm{dS}}}{\tau} = S_{\mathrm{dS}} T_{\mathrm{dS}} H
		= \frac{c^5}{2G}
	\end{equation}
	so given the assumptions, the lower power bound is half the Planck power, at $\sim 1.8 \times 10^{52} \ \text{W}$. This is an extremal bound under full saturation hypothesis.
	Finally, in terms of power density
	\begin{equation}
			\frac{	\dot Q_{\min}^{\mathrm{Tot}}}{V_\mathrm{dS}} \bigg|_{N=S_{\mathrm{dS}}/(k_B \ln 2)} 
			\ge\ \frac{S_{\mathrm{dS}} T_{\mathrm{dS}}}{V_\mathrm{dS} \tau} = \rho_\Lambda H 
	\end{equation}
	which, if evaluated today, has an order of magnitude  $\sim 10^{-27}\ \text{W/m}^3$.
	
	\paragraph*{Takeaway.} Far from being a claim on the origin of dark energy, this operational repackaging shows an alternative interpretation as a minimum informational cost. This is valid only under the assumptions of this thought experiment (bound saturation, unbiased record and, for the power, $\tau \sim H^{-1}$). Also, as maximum entropy $H_2 = \ln 2$ is also taken into account, the test can be any binary test, not necessarily a vacuum test.
	
	\section{Conclusions}
	
	In this work we framed a continuous vacuum/no-vacuum monitoring task as an \emph{operational} information-processing problem: at each coarse-grained time bin of duration \(\tau\), the measurement apparatus produces a binary classical record (click/no-click) that must ultimately be stored and, if one insists on a stationary operation, reset. Under this minimal and model-agnostic assumption, Landauer's principle provides a lower bound on the average heat dissipated per bin that is set by the Shannon entropy of the record.
	
	For a single monitored mode, modelling the test as a Poissonian click process with effective strength \(\gamma\), we obtained the bound
	\[
	\dot Q_{\min}^{(1)}(t)\ \ge\ \frac{k_B T}{\tau}\,
	H_2\!\left(1-e^{-\gamma(1-p_0(t))\tau}\right),
	\]
	highlighting a key point: the thermodynamic cost is not tied to the mode energy itself, but to the \emph{information throughput} of the classical outcomes that are being maintained and erased. In the near-vacuum regime (\(p_0\simeq 1\)), the click probability per bin is strongly biased, the record is highly compressible, and the optimal dissipation can be parametrically below the worst-case \(k_B T\ln 2\) per bin. Conversely, if one pushes the time resolution \(\tau\to 0\) at fixed instantaneous click rate, the corresponding \emph{power} bound grows without limit but only logarithmically in \(1/\tau\), making explicit the necessity of an operational cutoff such as finite detector bandwidth or coarse-graining time.
	
	Extending to many modes, we emphasized that the relevant quantity is the entropy of the \emph{joint} record \(H(\mathbf{Y}_t)\), which reduces to \(N H_2(P_\tau)\) only in the independent case. This clarifies how correlations and redundancies across modes can lower the minimal erasure cost through compressibility, while any attempt to attribute a cost density to a spatial region necessarily requires a finite mode density \(\nu=N/V\) fixed by a spatial and/or spectral cutoff.
	
	Finally, we discussed a speculative cosmological thought experiment in which a horizon-like system plays the role of the ultimate memory and thermal reservoir. In that language, the Landauer cost becomes \(k_B T_{\mathrm{dS}}\,H(\mathbf{Y}_t)\) per bin, and under strong saturation assumptions one can formally relate the maximal unbiased information throughput to characteristic de Sitter scales. We stress, however, that this does not constitute a derivation of dark energy: it is an operational re-interpretation that holds only within the assumptions made (unbiased records, saturation of the entropy budget, and a horizon-scale coarse-graining \(\tau\sim H^{-1}\)).
	
	Overall, the main message is that continuous “vacuum tests” (and, more generally, any binary monitoring protocol) have an unavoidable thermodynamic footprint once their classical record is treated as a physical resource. The magnitude and scaling of this footprint are controlled by detector coarse-graining, by the bias/compressibility of outcomes, and by the effective number of monitored degrees of freedom. These considerations provide a concrete bridge between measurement theory, information theory, and thermodynamic costs in continuous monitoring scenarios.

\end{document}